\documentclass[11pt,preprint]{aastex}

\def\ltsima{$\; \buildrel < \over \sim \;$}
\def\lsim{\lower.5ex\hbox{\ltsima}}
\def\gtsima{$\; \buildrel > \over \sim \;$}
\def\gsim{\lower.5ex\hbox{\gtsima}}

\slugcomment{}

\shorttitle{Limits on eclipses of KH~15D} 
\shortauthors{Winn, Garnavich, Stanek, \& Sasselov}

\begin{document}

\title{ Limits on eclipses of the pre-main-sequence star KH~15D\\ in
the first half of the 20th century }

\author{
Joshua N.\ Winn\altaffilmark{1,2},
Peter M.\ Garnavich\altaffilmark{3},
K.~Z.\ Stanek\altaffilmark{1},
Dimitar D.\ Sasselov\altaffilmark{1}
}

\email{jwinn@cfa.harvard.edu; pgarnavi@miranda.phys.nd.edu;
kstanek@cfa.harvard.edu; sasselov@cfa.harvard.edu}

\altaffiltext{1}{Harvard-Smithsonian Center for Astrophysics, 60
Garden st., Cambridge, MA 02138}

\altaffiltext{2}{National Science Foundation Astronomy \& Astrophysics
Postdoctoral Fellow}

\altaffiltext{3}{University of Notre Dame, Notre Dame, IN 46556}

\begin{abstract}
Over the last decade, the pre-main sequence star KH~15D has exhibited
periodic eclipses that are surprisingly deep ($\sim$3~mag) and
long-lasting ($\sim$40\% of the 48.4-day period). The cause of the
eclipses is unknown, but it could be a feature in a nearly edge-on
protoplanetary disk.  Here we report on an analysis of archival
photographs of KH~15D from the Harvard College Observatory plate
collection, most of which were taken during the years
1913-1951. During this time range, the data are consistent with no
eclipses; the duty cycle of 1~mag eclipses was less than 20\%. The
decadal timescale of this change in eclipse behavior is compatible
with the expected timescale of protoplanet/disk interactions. Archival
images from more recent epochs should reveal the onset of the
eclipses.
\end{abstract}

\keywords{ stars: pre-main sequence --- stars: individual (KH~15D) ---
open clusters and associations: individual (NGC~2264) --- planetary
systems: protoplanetary disks }

\section{Introduction}
\label{sec:intro}

The unique eclipses of the pre-main sequence star KH~15D have recently
gained attention because of the possibility that they are related to
processes of planet formation.  The eclipses were discovered by
\citet{kh98}\footnote{The name KH~15D was derived from their initials
and the star's catalog number in field D of \citet{keh+97}.} during a
variability study of the young cluster NGC~2264. The optical spectrum
of KH~15D is that of a T Tauri star \citep{hhs+01,hhm+03}, a category
believed to represent a late stage in the accretion and dispersal of
circumstellar material and the condensation of planets \citep{b89}.
\citet{hhs+01} estimated the age of the star to be 2--10~Myr, from its
position on a color-magnitude diagram.

During the eclipses, which recur every 48.36 days and last $\approx$20
days, the star plummets in brightness by over 3 magnitudes in a manner
consistent with a knife-edge crossing the face of the star (see Herbst
et al.\ 2002 for a striking presentation and discussion of the light
curves). The obscuring matter must be composed of large dust grains or
macroscopic objects, because small dust grains would cause reddening
of the starlight that is not observed. If the obscuring material is
orbiting the star, it must be spread out over a large fraction of the
orbit in order to explain the long duration of the
eclipses. Strangely, the star re-brightens near mideclipse, as if the
obscuring material is distributed symmetrically about a central
opening.

The eclipses are sufficiently dramatic that it appeared possible to
investigate their history in photographic plate archives, despite the
poor sensitivity and resolution of old photographs compared to modern
detectors. In \S~\ref{sec:data}, we describe our findings from the
Harvard College Observatory plate collection. In \S~\ref{sec:analysis}
we use the data to place limits on the eclipse properties of KH~15D in
the first half of the 20th century. Finally, in
\S~\ref{sec:discussion}, we place our results in the context of
previous measurements and previously offered theories for the origin
of the eclipses.

\section{Data}
\label{sec:data}

We searched the collection of the Harvard College Observatory, one of
the world's largest archives \citep{h94}, for plates of NGC~2264. The
most useful part of the archive was the ``MC series,'' one of the
highest-quality plate series in the collection.  It was obtained with
the 16-inch $f/5.2$ Metcalf Doublet refracting telescope, in Cambridge
and Oak Ridge, Massachusetts, between 1909 and 1988.  We found 66
suitable plates in the MC series.  In each case we digitized a
$30\arcmin\times 30\arcmin$ region surrounding the expected position
of KH~15D, using an optical scanner with a resolution of 3000
pixels~inch$^{-1}$ and 14 bits~pixel$^{-1}$.

We also found one suitable plate in the ``A series,'' taken with the
24-inch $f/5.6$ Bruce Doublet in Bloemfontein, South Africa, on 1936
January 24. This plate could not be digitized due to its large
physical size ($14\times 17$ inches). However, the A plate was of much
higher quality than any of the MC plates, allowing the analysis
described below to be performed visually.

To determine the threshold of detectability of each plate, we chose 10
stars with a range in brightness of roughly 2 magnitudes bracketing
the magnitude of KH~15D in its uneclipsed state, and searched for
those 10 stars in each image. These reference stars are identified by
KH number in Fig.~1a, which is an image of the field from the Palomar
Observatory Sky Survey (POSS)\footnote{The National Geographic
Society---Palomar Observatory Sky Atlas (POSS-I) was made by the
California Institute of Technology with grants from the National
Geographic Society.}.

Next, we looked for KH~15D. This was not a trivial task because its
uneclipsed magnitude was usually close to the limiting magnitude. In
addition, a nearby bright star (HD~47887) was often overexposed to
such a degree that its ``halo'' of scattered light enveloped the
expected position of KH~15D. This was a particular problem in
blue-light observations (which formed the majority of cases) because
the bright star is very blue. In those cases, we employed the
technique of ``unsharp masking'' commonly used in astrophotography
\citep[see, e.g.,][]{mz79}. In this technique, one produces a blurred
image by convolving the original with a circular Gaussian tapering
function, and then subtracts a scaled version of the blurred image
from the original image. This enhances contrast by removing low
spatial frequencies, at the expense of an increased noise level. The
width of the Gaussian function, and the scaling of the blurred image,
are adjusted to achieve optimal results. Here the method works because
the image of HD~47887 is very broad and flat-topped, whereas images of
fainter stars such as KH~15D are more sharply peaked (see Figs.~1b and
1c). Star KH~24D, being brighter than KH~15D and slightly further away
from HD~47887, provided a useful test of our ability to detect stars
close to the bright central star: if KH~24D could not be detected,
surely no useful information about KH~15D could be obtained from that
plate.

In this manner we determined that 40 of the plates had sufficient
sensitivity and resolution for KH~15D to have been detected when
uneclipsed. Among these, we securely detected KH~15D in 36 cases. In
the other 4 cases, we believe KH~15D is present, but we are not as
confident in the detection because it is right at the plate limit
(i.e.\ brighter comparison stars were detected, but fainter comparison
stars were not detected). We summarize the results as follows: in 90\%
of the cases for which detection of KH~15D was possible, it was
securely detected, and there were no secure non-detections.

We have not attempted accurate photometry because of the contaminating
influence of HD~47887. However, judging from the comparison stars, we
conclude that the detection of KH~15D on a given plate rules out an
eclipse deeper than 1~mag at that epoch.  Thus, the data are
consistent with no eclipses deeper than 1~mag.

\section{Analysis}
\label{sec:analysis}

Fig.~2 shows the time coverage of the 40 high-quality plates, and also
includes the POSS image (1951 December 1) and the 2 epochs of the
DPOSS-2 images\footnote{The Digitized Sky Surveys were produced at the
Space Telescope Science Institute under U.S. Government grant NAG
W-2166. The images of these surveys are based on photographic data
obtained using the Oschin Schmidt Telescope on Palomar Mountain and
the UK Schmidt Telescope. The plates were processed into the present
compressed digital form with the permission of these institutions.}
(1989 November 7 and 1992 December 21), on which KH~15D appears
uneclipsed. The data fall naturally into four groups in time, which we
have distinguished by arbitrary colors and symbols: black triangles,
for 1913--1915; red circles, for 1924--1939; green squares, for
1946--1955; and blue diamonds, for 1978--1992. Solid symbols are
secure detections and open symbols are the ambiguous cases.

We use this data to place limits on the ``eclipse fraction'' $f$,
defined as the fraction of time that KH~15D is eclipsed by
$>$1~mag. The simplest approach is to ignore the modern ephemeris and
assume that the phase of each epoch is random. We also assume,
conservatively, that the 4 lower-confidence detections represent
eclipses. Then, using binomial statistics with a uniform prior on $f$,
the observation that KH~15D is uneclipsed in 39/43 cases implies
$f<0.20$ with 95\% confidence. The result changes to $f<0.24$ when
only the 1924--1939 data are used, and $f<0.17$ when only the
1946--1955 data are used.

A more sophisticated approach is to use the modern ephemeris to
convert the time of each observation into $\phi$, the phase of the
photometric period. \citet{hhs+01} estimated the period to be
$48.34\pm 0.02$ days. After obtaining additional data, \citet{hhv+02}
revised the period to 48.36 days but did not give an error estimate.
We adopt the ephemeris of \citet{hhv+02},
\begin{equation}
{\rm JD (mideclipse)} = 2,452,352.26 + 48.36E,
\end{equation}
and an uncertainty of 0.02~days in the period. The phase coverage of
our observations is shown in Fig.~3, for the choices $P=48.34$, 48.36,
and 48.38 days. The dashed lines show the modern eclipse duration.
The dotted lines show the approximate phases of the mideclipse
re-brightenings, during which the star has sometimes been observed at
its uneclipsed brightness (or even slightly brighter).

The phase coverage is fairly complete and can be used to rule out
eclipses with the present characteristics. Suppose, for example, that
the past eclipses were nearly symmetric about mideclipse, as they are
today. We define the ``eclipse duty cycle'' $f'$ as the fraction of
the photometric period between the 1~mag levels of ingress and
egress. In general, $f'\geq f$ because of the possibility of
mideclipse re-brightenings. We determine the maximum phase of the
eclipses, $\phi_{\rm max}$, as the minimum $|\phi|$ among all the
secure detections, not counting detections with $|\phi| < 0.05$ (where
re-brightenings are expected). The resulting limit on the eclipse duty
cycle is $f'< 2\phi_{\rm max}$. To deal with the period uncertainty,
we determine $\phi_{\rm max}$ for all possible values of $P$, and then
compute a weighted histogram of the results, using a weighting
function that is Gaussian in $P$ with mean 48.36~days and
$\sigma=0.02$~days.

Using all the data, the result is $f'<0.14$ with 95\% confidence.
Using only the data from 1924--1939, the result is $f'<0.18$.  With
only the 1946--1955 data, the result is $f'<0.24$.  Neither the
earliest data nor the most recent data are constraining by themselves.

\section{Summary and discussion}
\label{sec:discussion}

We conclude that the deep eclipses of KH~15D observed today did not
occur in the first half of the 20th century with their present
characteristics.  The eclipses may have been shorter in duration or
shallower than 1~mag, or perhaps the mideclipse re-brightenings
extended to larger phases. Of course, a combination of these effects
or a more complex time history are also possible.

The simplest explanation is that the eclipses have grown in duration,
as has been observed over the last decade.  Fig.~4 shows the upper
limits on the eclipse duty cycle derived at the end of the previous
section, along with modern values that we estimated from the light
curves of \citet{hhv+02} and from unpublished data from 2002-3
obtained by a group led by one of the authors (P.G.).  Apparently a
change is occurring on decadal timescales, in either the features of
the obscuring material surrounding the KH~15D system, or in the
alignment of those features with our viewing angle.

What could explain both the eclipses and their recent onset?
\citet{hhs+01} and \citet{hhv+02} offered two hypotheses to explain
the eclipses: (1) KH~15D has a nearly edge-on protoplanetary disk, and
a warp or ridge in the disk periodically blocks the star. (2) KH~15D
has a higher-mass companion that is hidden by an edge-on circumstellar
disk. The observed star is eclipsed as its orbit carries it through
the projected plane of the disk. \citet{gt02} offered another
hypothesis: (3) KH~15D has a low-mass companion, and the resulting
accretion and outflow of circumstellar material creates an asymmetric
common envelope, with a large and dense region near the low-mass
companion. The orbital motion of the secondary causes periodic
occultations of the primary.

Hypothesis (2) seemed unlikely even before this work, because
\citet{hhv+02} and \citet{hhm+03} measured only a small radial
velocity shift ($3.3\pm 0.6$~km~s$^{-1}$) between widely separated
phases. Those authors also expressed concern about systematic effects
that may allow the true radial velocity shift to be zero. In addition,
we do not know how to make this hypothesis compatible with the recent
onset of the eclipses.

The low-mass companion of hypothesis (3) could be small enough to be
allowed by the current radial velocity measurements. \citet{gt02} do
not make any specific statements regarding the expected timescale of
changes in eclipse duration, but they do claim that their model can
explain decadal variations seen in other young photometrically active
stars such as UX Ori stars. This makes hypothesis (3) viable and
worthy of further investigation.

Hypothesis (1) is not only consistent with a small radial velocity
shift, but also naturally involves a physical timescale that is
compatible with our result. The viscous timescale at the orbital
distance of 0.2~A.U.\ (corresponding to the 48-day period) is 10--100
years. If protoplanet/disk interactions cause perturbations in the
disk, then one would expect these perturbations to evolve on this
timescale. All other timescales in the disk at 0.2~AU are much
shorter: the dynamical and cooling timescales are $\sim$2 days, while
a single-impulse perturbation would be damped in less than 1 month
\citep{dch+99}. Thus, protoplanet/disk interactions are an appealing
explanation of the KH~15D phenomenon and the observed evolution.

For the sake of providing a concrete and falsifiable example of such
an interaction, we indulge in further speculation, elaborating on a
suggestion by \citet{hhv+02}. Models of planet formation at larger
orbital distances predict density perturbations in the form of two
spiral waves: a leading wave extending to smaller orbital distances,
and a lagging wave extending to larger distances \citep[see,
e.g.,][]{blo+03}. The spiral waves extend more than a radian in
azimuth, and for a Jupiter-mass planet the density is enhanced by a
factor of 3-5 even at one pressure scale-height, $h$, above the
midplane. For a typical T Tauri disk, the optically thick portion of
the disk extends to a height of $H_{s}\approx 4h$ above the midplane
\citep{dch+99,js03}, corresponding to 0.02~A.U.\ at the estimated
orbital distance (0.2~A.U.) of the occulting feature. This thickness
is approximately equal to the diameter of the young star ($2R_s
\approx 4R_{\odot} = 0.02$~A.U.). The spiral shock waves induced by a
Jupiter-sized protoplanet are capable of increasing $H_{s}$ by more
than a factor of 2 and creating an elongated ridge along our line of
sight that is thick enough to block the central star
entirely. Moreover, as previously suggested \citep{hhv+02} and
recently demonstrated with three-dimensional computational models
\citep{blo+03}, the ridge will have a local depression at the position
of the protoplanet, which would explain the mid-eclipse
re-brightenings.

If this is correct, our na\"{\i}ve prediction is that the eclipse duty
cycle will increase to $\approx$0.75 and remain there for the next
decade, because in the simulations, the spiral waves cover a maximum
of $\approx$75\% of the orbit. Furthermore, if the true period is
actually 97 days, as suggested by \citet{hhv+02} based on small
differences between alternate eclipses, this scenario would be
wrong. We emphasize that this speculation was inspired by simulations
at much larger orbital distances; we are not aware of any detailed
simulations of protoplanet/disk interactions at 0.2~A.U.

Finally, we note that although the Harvard collection does not happen
to include many photographs of this field after 1960, many such
photographs undoubtedly exist. Because of its placement in a
scientifically interesting young cluster, and near a photogenic
nebula, there should be many images from 1960 onward that are of
sufficient quality to measure KH~15D.  A compilation of these data
should reveal the onset of the modern eclipse behavior.

\acknowledgments The authors wish to thank Alison Doane and Doug Mink
for valuable assistance with the plate archive. This material is based
upon work supported by the National Science Foundation under Grant
No.\ 0104347.

\newpage

\begin{figure}
\epsscale{1.0}
\plotone{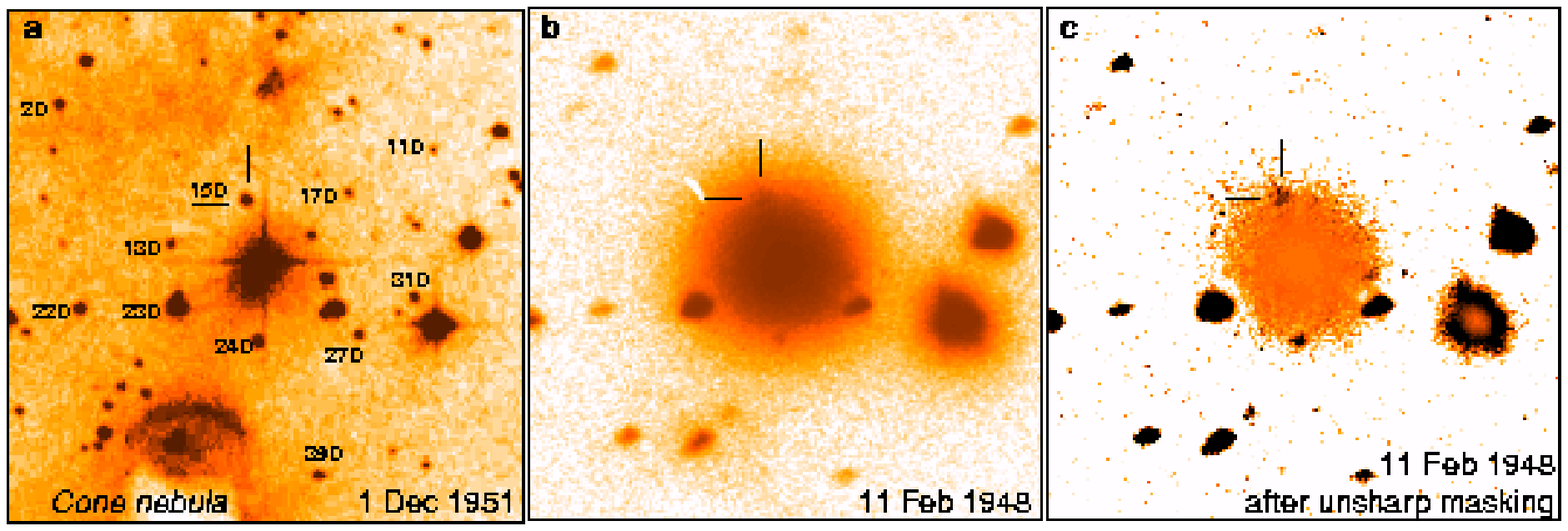}
\caption{ The 4.8 arcminute region surrounding HD~47887 from (a) the
POSS, (b) a typical blue plate, and (c) the same plate after unsharp
masking. In all cases, north is up and east is left. }
\label{fig1}
\end{figure}

\begin{figure}
\epsscale{1.0}
\plotone{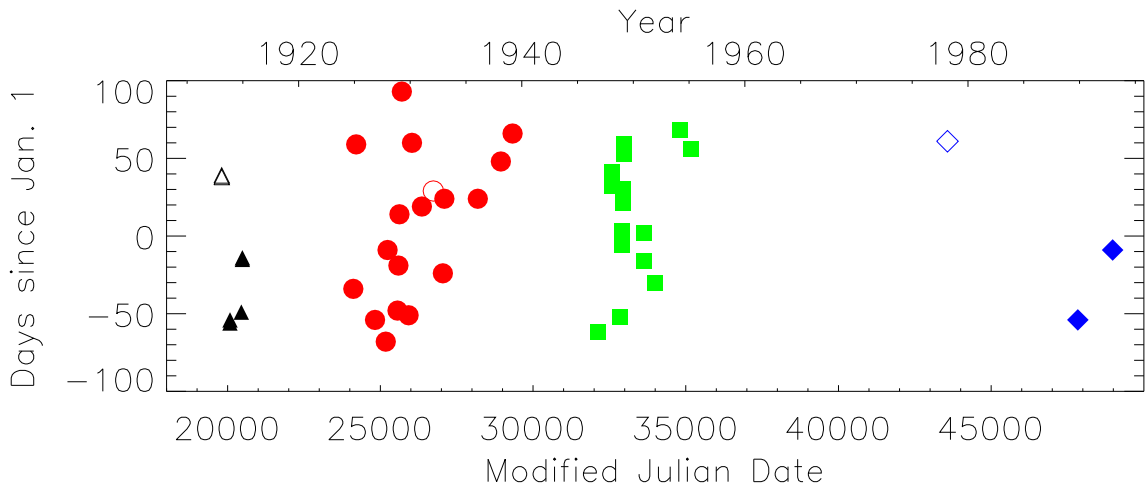}
\caption{ Time coverage of 40 high-quality plates and also the DPOSS
and DPOSS-2 images. In order to separate the points vertically, the
$y$-axis was taken to be the number of days since the nearest
Jan.~1. }
\label{fig2}
\end{figure}

\begin{figure}
\epsscale{0.7}
\plotone{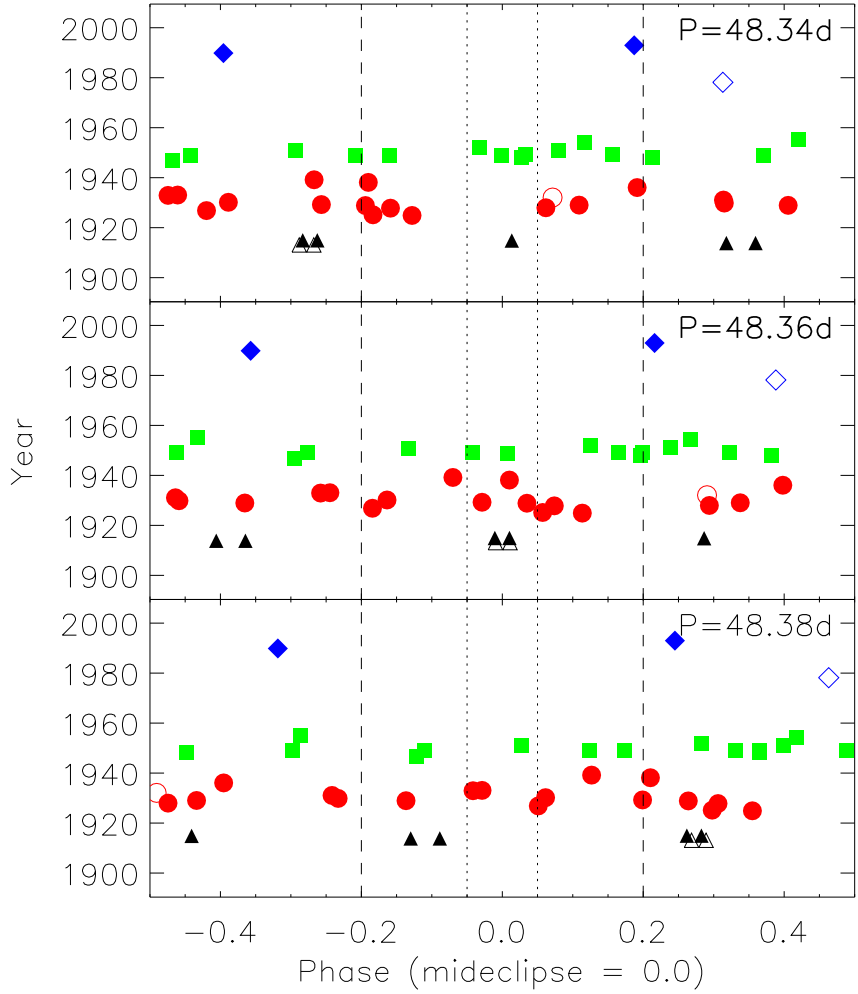}
\caption{ Phase coverage of detections of KH~15D, for three choices of
period $P$. The symbols and color scheme are the same as in Fig.~2. }
\label{fig3}
\end{figure}

\begin{figure}[ht]
\epsscale{1.0}
\plotone{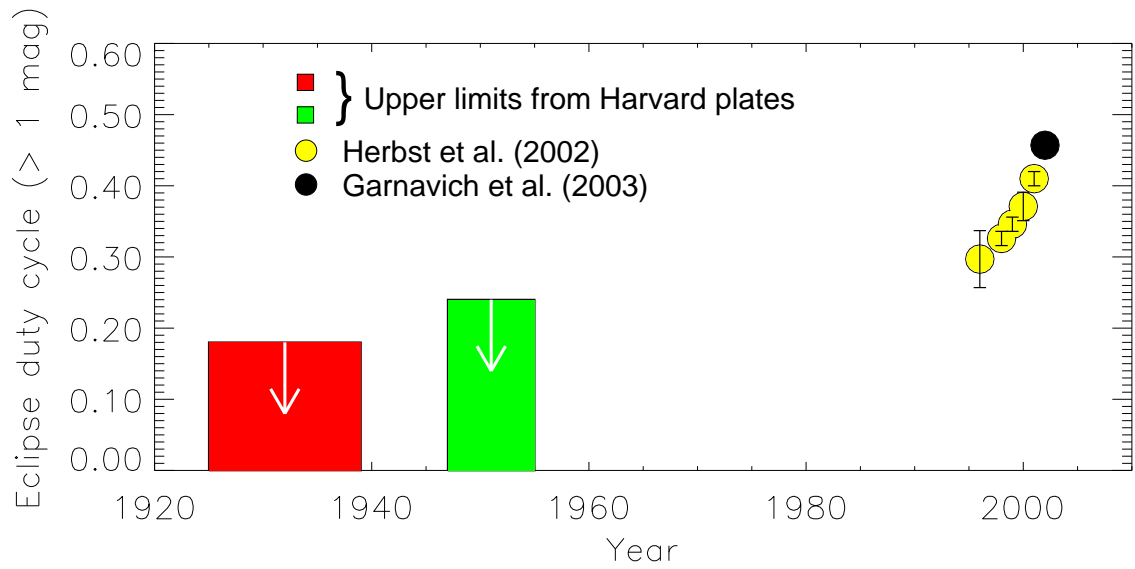}
\caption{ Determinations of the eclipse duty cycle ($f'$) from this
work, from our estimates using the light curves of Herbst et al.\
(2002), and from unpublished data by Garnavich et al.\ (2003). }
\label{fig4}
\end{figure}


\begin{thebibliography}{}

\bibitem[Bate et al.(2003)]{blo+03} Bate, M.R., Lubow, S.H., Ogilvie,
G.I., \& Miller, K.A.\ 2003, \mnras, 341, 213

\bibitem[Bertout(1989)]{b89} Bertout, C.\ 1989, \araa, 27, 351

\bibitem[D'Alessio et al.(1999)]{dch+99} D'Alessio, P., Cant\'{o}, J.,
Hartmann, L., Calvet, N., \& Lizano, S.\ 1999, \apj, 511, 896

\bibitem[Grinin \& Tambovtseva(2002)]{gt02} Grinin, V.~P.\ \&
Tambovtseva, L.~V.\ 2002, Ast.\ Lett.\ 28, 601

\bibitem[Hamilton et al.(2001)]{hhs+01} Hamilton, C.M., Herbst, W.,
Shih, C., \& Ferro, A.J.\ 2001, \apj, 554, L201

\bibitem[Hamilton et al.(2003)]{hhm+03} Hamilton, C.~M., Herbst, W.,
Mundt, R., Bailer-Jones, C.~A.~L., \& Johns-Krull, C.~M.\ 2003,
preprint [astro-ph/0305477]

\bibitem[Hazen(1994)]{h94} Hazen, M.~L.\ 1994, IAU Symp.~161:
Astronomy from Wide-Field Imaging, 161, 365

\bibitem[Herbst et~al.(2002)]{hhv+02} Herbst, W.\ et al.\ 2002, \pasp,
114, 1167

\bibitem[Jang-Condell \& Sasselov(2003)]{js03} Jang-Condell, H.\ \&
Sasselov, D.\ 2003, \apj, in press [astro-ph/0304330]

\bibitem[Kearns et al.(1997)]{keh+97} Kearns, K.~E., Eaton, N.~L.,
Herbst, W., \& Mazzurco, C.~J.\ 1997, \aj, 114, 1098

\bibitem[Kearns \& Herbst(1998)]{kh98} Kearns, K.~E.\ \& Herbst,
W.~H.\ 1998, \aj, 116, 261

\bibitem[Malin \& Zealey(1979)]{mz79} Malin, D.~F.~\& Zealey, W.~J.\
1979, Sky \& Telescope, 57, 354

\end{thebibliography}
\end{document}